\documentclass[preprint,showpacs,nofootinbib,prd,aps]{revtex4-1}
\usepackage{graphicx,amstext,amssymb}

\begin{document}
\title{Hot origin of the Little Bang}

\author{S.V. Akkelin$^{1,2,3}$}
\affiliation{$^1$Bogolyubov Institute for Theoretical Physics,
Metrolohichna  14b, 03680 Kiev,  Ukraine\\
 $^2$ Institut f\"ur Theoretische Physik, Universit\"at Heidelberg,
Philosophenweg 16, 69120 Heidelberg, Germany\\
$^3$ ExtreMe Matter Institute EMMI, GSI Helmholtzzentrum f\"ur
Schwerionenforschung, Planckstra\ss e 1, 64291 Darmstadt, Germany }

\begin{abstract}

Ultrarelativistic heavy ion collisions produce a quark-gluon matter
which lies in the future light cone originating from given points on
the $t=z=0$ plane of the Minkowski spacetime manifold. We show that
in a weak coupling regime the Minkowski vacuum of  massless fields
presents itself in the ``Little Bang'' region  as a thermal state of
low $p_{T}$ particles, in close analogy to the Unruh effect for
uniformly accelerated observers which are causally restricted to a
Rindler wedge.  It can shed some light on the mechanisms of early
time thermalization  in ultrarelativistic heavy ion collisions.

\end{abstract}

\pacs{25.75.-q,03.70.+k}

 \maketitle

 \section{Introduction}

The   ``Little Bang'' made in an ultrarelativistic heavy ion
collision of the two Lorentz contracted nuclei starts  from the
spacetime region near $t \simeq z \simeq 0$ (in the laboratory
frame),  and produces initially very dense quark-gluon matter which
then expands (mostly in the longitudinal direction) and eventually
undergoes a transition to hadronic degrees of freedom (for recent
reviews see, e.g., Refs. \cite{Gelis-a,Gelis-b,PBM} and references
therein). It is firmly established now  that the quark-gluon matter
created in these collisions seems to behave like a fluid
\cite{Hydro-a,Hydro-b,Hydro-c,Hydro-d} on very short time scales,
but a first principle proof of the applicability of hydrodynamics
for the initially far-from-equilibrium quark-gluon plasma is still
lacking (for first-principles description of the early-time dynamics
in ultrarelativistic heavy ion collisions see, e.g., Refs.
\cite{Berges-a,Berges-b}) even though considerable progress has been
made in the past years (for recent reviews see, e.g., Refs.
\cite{Kurkela-a,Kurkela-b,P} and references therein). There are many
similarities (as well as differences) between the physics of  Little
Bang fireballs created in ultrarelativistic heavy ion collisions,
and the cosmology (see, e.g., Refs. \cite{Bang-a,Bang-b}). Even
widely utilized $t,z$ parametrization of spacetime region occupied
by the expanded fireball \cite{Bjorken} is nothing but the Milne
coordinates \cite{Milne,Jeon} which  are, in fact, merely
two-dimensional coordinate transformations of corresponding  (future
light cone) causally-connected region of the flat (Minkowski)
spacetime manifold.

If one take such a  Milne ``universe'' (future light cone) as a
spacetime in its own right, then one can define vacuum state (or
states) in the corresponding spacetime region. Indeed, it is well
known (see, e.g., Ref. \cite{Milne}) that solution of the free
Klein-Gordon equation for massive scalar field in the
two-dimensional Milne universe can be written in terms of either
Bessel or Hankel functions, from which two complete sets of
normalized modes, that are related by the Bogolyubov transformation,
can be constructed. One of the pure vacuum states (associated with
the Hankel functions) is, in fact, analogous  to the usual Minkowski
vacuum which is the ground state with respect to the Hamiltonian
that is generator of the time translations. This vacuum is
ill-defined in the massless limit for zero transverse momenta (then
$m_{T}=\sqrt{m^{2}+p_{T}^{2}}=0$), as well as initially, at $t=0$,
for any $m_{T}$. The other one (associated with the Bessel
functions) is analogous to the Rindler vacuum of an uniformly
accelerated observer. There is thermal-like relation between these
vacuum states, and the corresponding temperature is inversely
proportional to the time, see details in Ref. \cite{Milne}.

In the present article we take a different approach. Namely, we
assume  that unlike true Milne universe, which is a spacetime in its
own right, the spacetime region occupied by the matter produced in
an ultrarelativistic heavy ion collision is embedded into the larger
Minkowski spacetime with corresponding (Minkowski) vacuum state.
Then the question arises: how does a ``relevant'' restriction of the
Minkowski vacuum state to the Milne spacetime subspace look like for
corresponding (local) operators which belong to this subspace
?\footnote{Notice that unlike the classical field theory, where
vacuum really means an empty space, the quantum field vacuum is a
pure state which contains nontrivial space-like quantum
correlations. The reason is the following: while the commutator of
field operators does vanish when $x$ and $y$ are space-like related,
the expectation value $<0|\phi(x)\phi(y)+ \phi(y)\phi(x)|0> $ is not
equal to $2<0|\phi(x)|0><0|\phi(y)|0>$, it is a manifestation of
space-like quantum correlations (entanglement) in the vacuum. }

The problem of how the Minkowski vacuum state looks like for an
observer restricted to some causally-connected spacetime region is
not new and was discussed earlier. A well known example is the Unruh
effect \cite{Unruh-0} (for reviews see, e.g., Refs.
\cite{Milne,Unruh-rev-0,Unruh-rev}, and for discussions about the
validity of the Unruh effect see Refs. \cite{d-1,d-2,d-3}) which is
the result of a restriction of an uniformly accelerated observer to
a Rindler wedge whose  borderlines correspond to the event horizon.
It was shown that the Minkowski vacuum expectation values of the
local field operators restricted to this wedge  seem to be
calculated in an impure thermal-like state with respect to the wedge
preserving generator of the corresponding time-like translations,
and this generator is proportional to a Hamiltonian that is
associated to a proper time of an uniformly accelerated (and,
thereby, ``eternal'') observer.\footnote{For analysis  of cases
where an observer undergoes non-uniform acceleration  see e.g. Refs.
\cite{nonuniform-1,nonuniform-2}.} The corresponding ``temperature''
is proportional to an constant proper acceleration of an uniformly
accelerated observer. For free fields, it was demonstrated that the
Unruh effect follows from the fact that the Minkowski vacuum state
in the Rindler basis can be written as an entangled state between
two sets of modes, respectively spanning left and right Rindler
wedges \cite{Unruh-0,Unruh-1,Unruh-rev-0,Unruh-rev}. Then,
accounting that an uniformly accelerated observer is constrained to
move in one of the Rindler wedges, one gets that the  reduced
density matrix corresponds  to a thermal state with the Unruh
temperature proportional to the observer's acceleration.
Interestingly, a similar result was recently obtained in Ref.
\cite{Olson} for two-dimensional scalar field quantized in the Milne
coordinates. Namely, it was shown that between the massless free
fields within the future and past light cone there is the same
entanglement as for fields between the left and right Rindler
wedges,  and the existence of a vacuum thermal effect for an
inertial observer constrained to interact with the field in only the
future or the past light cone was noted. It is also worth noting
that in Ref. \cite{Rovelli} the Unruh effect was studied for an
observer with a finite lifetime, who has access to the local
observables associated to a finite spacetime region called a
``diamond''. Using the thermal time hypothesis, it was shown that
the Unruh effect exists for such an observer too, and that
corresponding temperature is time-dependent and does not vanish even
in the limit in which the acceleration is zero.

In our opinion, the possible existence  of  an analogue of the Unruh
effect in the Milne ``universe'' (future light cone) appears worthy
of further investigation, especially in respect to the longstanding
problem of early time thermalization in relativistic heavy ion
collisions, see e.g. Refs.
\cite{Gelis-a,Gelis-b,Kurkela-a,Kurkela-b}.\footnote{The Unruh
effect has been considered as a possible explanation of the observed
thermal behavior  in relativistic heavy ion collisions in Refs.
\cite{Kharzeev-1-a,Kharzeev-1-b,Kharzeev-1-c,Kharzeev-2-a,Kharzeev-2-b,Kharzeev-2-c,Kharzeev-2-d},
where the dynamical origin of the (transient)  acceleration is
related with the strength of the color field.} In the present
article, we demonstrate that in a weak coupling approximation of
effectively two-dimensional massless scalar field model the
Minkowski vacuum state looks like the mixed thermal state  for
operators restricted to the future light cone with the corresponding
time-like generator as the Hamiltonian, an analogue to the Unruh
effect. We obtain our results using the method which was early
applied in Ref. \cite{Unruh-2} to derive the Unruh effect for scalar
field with interactions.

\section{Minkowski vacuum  and relativistic heavy ion collisions}

Most of the particles detected in an ultrarelativistic heavy ion
central collision are produced by the relevant subsystem which is
created in the future light cone with the beginning at the $t=z=0$
plane (more exactly, at spacetime region near $t \simeq z \simeq
0$), where two Lorentz-contracted nuclei collide and interact. The
(relevant) past evolution of this subsystem  is encoded into the
corresponding initial state within  the future light cone. If one
defines an initial state at $\tau = \sqrt{t^{2}-z^{2}}=const$
hypersurface, then such a hypersurface encloses a spacetime region
within the future light cone. Separation of such a subsystem means
that one needs trace over unobservables (e.g., correlations with
irrelevant degrees of freedom outside the future light cone).
Similar to quantum mechanics, such a trace-out procedure can result
in loss of an information. It is worth noting  that quantum field
theory is in a certain sense more ``quantum'' than quantum mechanics
because presence of  the quantum vacuum in the former. In quantum
field theory, unlike of quantum mechanics, there are vacuum
correlations (entanglements) between quantum fields which are
localized inside the region enclosed by the hypersurface
$\tau=const$, and quantum fields localized outside the one.
Therefore, one can expect that under certain conditions the
Minkowski vacuum can look like (time-dependent) thermal mixed state
for quantum operators depending on the values of quantum field in
spacetime points belonging only the region enclosed by the
$\tau=const$ hypersurface. In such a case the expectation values of
corresponding operators in the Minkowski vacuum state can coincide
with ones calculated in some thermal-like state just due to the
entanglement of quantum vacuum fluctuations.

To demonstrate how this idea works, let us consider a massless
scalar quantum field model  with a classical action
\begin{eqnarray}
S=\int dt d^{3}r \left [\frac{1}{2} \left (\frac{\partial
\phi}{\partial t}\right )^{2} - \frac{1}{2} \left (\frac{\partial
\phi}{\partial \textbf{r}}\right )^{2} - V(\phi) \right ] = \int dt
d^{3}r L, \label{1}
\end{eqnarray}
where $\textbf{r}=(x,y,z)$, $V(\phi)$ is a polynomial function of
$\phi$, and $L$ is the corresponding Lagrangian density in the
global Minkowski spacetime.

It is well known that in the path-integral formulation of quantum
field theory the transition amplitude from the initial vacuum $|
0,in \rangle $ to the final vacuum $| 0,out \rangle $ in the
presence of a source $J$ is given by the generating functional ($N$
is a normalization factor)
\begin{eqnarray}
Z[J]=\langle 0,out | | 0,in \rangle = N \int D \phi \exp{i(S(\phi)+
J \phi )}, \label{01}
\end{eqnarray}
and we use the shorthand symbol $J  \phi $ for four-dimensional
integral over product of external source and $\phi$. The functional
integration is taken over the space of all possible functional forms
of $\phi$ with some initial and final boundary values.

The expectation values of field operators in the Heisenberg picture
can be obtained by functional differentiations with respect to the
external classical source $J$. If $J=0$, these two vacua coincide
and reduce to the time-translation invariant vacuum $| 0 \rangle $.
Then initial and final boundary $\phi$ values coincide, and
\begin{eqnarray}
Z \equiv  Z[0] = \langle 0 | | 0 \rangle = \langle 0 |
e^{-iH^{[t]}(t_{out}-t_{in})}| 0 \rangle = N \int D \phi
\exp{(iS(\phi))} \label{02}
\end{eqnarray}
becomes the  vacuum to vacuum transition amplitude.  Here $H^{[t]}$
is the corresponding Hamiltonian, the superscript $[t]$ means that
the Hamiltonian $H^{[t]}$ is the generator of the time translation
in the flat Minkowski spacetime.

Let us recall that one can regard $Z$ as the partition function for
a thermal system at zero temperature (see, e.g., Refs.
\cite{Bernard,integral-a,integral-b}) with respect to the
Hamiltonian $H^{[t]}$. With this aim in view  let us assume that
$t_{out}= -t_{in}=\infty$. Then, by making the time pure imaginary
(so called Euclidian time), $it=t_{E}$, $t_{E}$ is real, we get the
Euclidian action
\begin{eqnarray}
S_{E} =\int dt_{E}  d^{3}r \left [\frac{1}{2} \left (\frac{\partial
\phi}{\partial t_{E}}\right )^{2} +\frac{1}{2} \left (\frac{\partial
\phi}{\partial \textbf{r}}\right )^{2}+ V(\phi) \right ] = -\int
dt_{E} d^{3}r L(\phi,i\frac{\partial \phi}{\partial t_{E}}
).\label{4}
\end{eqnarray}
Now, let integration goes over all periodic paths that have the same
classical $\phi$ values at $t_{E}=-\infty$ as at $t_{E}=\infty$.
Then  the Euclidian  functional,
\begin{eqnarray}
Z_{E}=N_{E}\int_{periodic} D \phi \exp{(-S_{E})}, \label{3}
\end{eqnarray}
where $N_{E}$ is a normalization factor,  is equivalent to the
thermal partition function $Tr[e^{- \beta H^{[t]}}] $ at zero
temperature, $T^{-1}=\beta \rightarrow \infty $. Indeed,  one can
see that
\begin{eqnarray}
Z_{E} = \lim_{\beta \rightarrow \infty } Tr[ | 0 \rangle \langle 0 |
e^{- \beta E_{0}}],  \label{04}
\end{eqnarray}
and if we assume that the energy of the Minkowski vacuum, $E_{0}$,
is zero (it is possible if the spectrum of the Hamiltonian is
bounded from below), we finally get
\begin{eqnarray}
Z_{E} = Z = \langle 0 | | 0 \rangle . \label{05}
\end{eqnarray}
This  allows one to consider the usual vacuum as a zero temperature
thermal state, and Eq. (\ref{3}) as the Euclidean functional
integral representation of the Minkowski vacuum.

It is important to note that a global vacuum state of the field at
zero temperature in the complete space can turn into a thermal state
with non-zero temperature in the incomplete space \cite{Unruh-2}. It
can easily be seen using polar variables,
\begin{eqnarray}
t_{E}=a^{-1}e^{a\eta}\sin a\zeta  , \label{5}\\
z =a^{-1}e^{a\eta}\cos a \zeta , \label{6}
\end{eqnarray}
where  $a$ is a scale-parameter, and performing change of
integration variables in Eq. (\ref{4}). Then we get
\begin{eqnarray}
S_{E} =  \int_{0}^{\beta_{R}} d \zeta \int d^{2}r_{T}  d \eta \left
[\frac{1}{2} \left (\frac{\partial \phi}{\partial \zeta}\right )^{2}
+ \frac{1}{2} \left (\frac{\partial \phi}{\partial \eta}\right )^{2}
+ e^{2a\eta}\left (\frac{1}{2}\left (\frac{\partial \phi}{\partial
\textbf{r}_{T}}\right )^{2}+ V(\phi)\right ) \right], \label{7}
\end{eqnarray}
where
\begin{eqnarray}
\beta_{R} = \frac{2 \pi }{a} \label{8}
\end{eqnarray}
in order to cover the whole Euclidian $(t_{E},\textbf{r}_{T},z)$
space, and we allow $\int D\phi$ to go over all periodic paths. One
can see that the Euclidian functional integral (\ref{3}), (\ref{7}),
(\ref{8}) is just a representation of the thermal partition function
with periodic boundary conditions $\phi(\zeta =0) = \phi(\zeta
=\beta_{R})$ with respect to imaginary ``time'' $\xi = i \zeta$,
$\zeta$ is real,
\begin{eqnarray}
Z=Tr[e^{- \beta_{R} H^{[\xi]}}], \label{9}
\end{eqnarray}
where the ``Hamiltonian'' $H^{[\xi]}$ is  the generator of
translations in the time-like direction with respect to real $\xi$.
Note that the thermal bath refers to the Hamiltonian $H^{[\xi]}$
which is different from the Hamiltonian $H^{[t]}$ whose lowest
energy eigenstate defines the Minkowski vacuum.

Now, let us recall how  the Unruh effect   is related with the above
formal approach (for details see Refs. \cite{Unruh-2,Unruh-rev}). We
begin with introducing the so-called Rindler coordinates in the
Minkowski spacetime. Namely, the flat spacetime expressed in the
globally defined Minkowski coordinates $(t,x,y,z)$ can be divided by
lines $t=\pm z$ into four quadrants which we call as future
($t>|z|$) and past ($-t>|z|$) light cones, and the right ($z> |t|$)
and left ($-z> |t|$) Rindler wedges. Rindler coordinates $(\xi,
\eta)$ are related with Minkowski coordinates $(t, z)$ in the right
Rindler wedge as
\begin{eqnarray}
z= a^{-1}e^{a\eta} \cosh a \xi, \label{10} \\
t=a^{-1}e^{a\eta}\sinh a \xi . \label{11}
\end{eqnarray}
Notice that the $(- \infty, + \infty)$ region of  $\xi$, $\eta$
coordinates covers   the right Rindler wedge, the two other
coordinates $\textbf{r}_{T}=(x,y)$ are the same both in the
Minkowski and Rindler frames. The trajectory defined by constant
values of $\eta$ and $\textbf{r}_{T}$ describes the motion of an
observer with constant proper acceleration
$ae^{-a\eta}>0$.\footnote{If $(\xi,\textbf{r}_{T},\eta)$ is treated
not as a physical frame of reference but  as an  abstract
non-inertial coordinate system, then such an  observer is an
hypothetical  one.} Using (\ref{10}), (\ref{11}), the Minkowski line
element restricted to the Rindler wedge becomes
\begin{eqnarray}
ds^{2}=dt^{2}-d\textbf{r}_{T}^{2}-dz^{2}=e^{2a\eta}(d \xi^{2}-
d\eta^{2})-d\textbf{r}_{T}^{2}. \label{12}
\end{eqnarray}
One can see that $e^{a\eta}\xi$ is a proper time of the uniformly
accelerated Rindler observer, and the Rindler time-like coordinate
$\xi$ is a measure of a proper time  along the trajectory with
$\eta=0$.

To define the Lagrangian density in the Rindler coordinates, notice
that the action (\ref{1}) for the scalar field with potential
$V(\phi)$ can be rewritten in the Rindler frame  (\ref{10}),
(\ref{11}) as
\begin{eqnarray}
S=\int d\xi d\eta  d^{2}r_{T} \left [\frac{1}{2} \left
(\frac{\partial \phi}{\partial \xi}\right )^{2} - \frac{1}{2} \left
(\frac{\partial \phi}{\partial \eta}\right )^{2} - e^{2a\eta}\left
(\frac{1}{2} \left (\frac{\partial
\phi}{\partial \textbf{r}_{T}}\right )^{2}+ V(\phi) \right ) \right]\equiv\\
\label{13.1}  \int d\xi d\eta  d^{2}r_{T} L_{R}, \label{13.2}
\end{eqnarray}
where $L_{R}$ is the corresponding Lagrangian density, and we take
into account that
\begin{eqnarray}
dtdz=d\eta d \xi e^{2a\eta}. \label{12.1}
\end{eqnarray}
The corresponding Hamiltonian density, $H_{R}$, then reads
\begin{eqnarray}
H_{R}=\Pi^{[\xi]}\frac{\partial \phi }{\partial \xi} -L_{R}=
\frac{1}{2} \left (\frac{\partial \phi}{\partial \xi}\right )^{2}+
\frac{1}{2} \left (\frac{\partial \phi}{\partial \eta}\right )^{2}+
e^{2a\eta}\left (\frac{1}{2} \left (\frac{\partial \phi}{\partial
\textbf{r}_{T}}\right )^{2}+ V(\phi) \right ), \label{14}
\end{eqnarray}
where $\Pi^{[\xi]}=\frac{\partial \phi}{\partial \xi}$ is the
conjugated field momentum.

The well known equivalence in the Rindler wedge between the
Minkowski vacuum partition function, $Z= \langle 0 | | 0 \rangle $,
and thermal partition function,
\begin{eqnarray}
Z_{R}=Tr[e^{- \beta_{R} H^{[\xi]}}], \label{14.1}
\end{eqnarray}
defined by the Rindler Hamiltonian $H^{[\xi]}=\int d\eta
d^{2}r_{T}H_{R}$ which generates  evolution along $\xi$   in
Minkowski spacetime, can easily be seen using an Euclidean
functional integral representation of the thermal partition function
(\ref{14.1}):
\begin{eqnarray}
Z_{R}=N_{E} \int D \phi \exp{\left [   \int_{0}^{\beta_{R}} d \zeta
\int d^{2}r_{T}  d \eta L_{R} (\phi, i\frac{\partial \phi}{\partial
\zeta} )\right ]}, \label{15}
\end{eqnarray}
with periodic boundary conditions. Now,  to derive the Unruh effect,
notice that substitution of the imaginary ``time'' $\xi = i \zeta$,
$\zeta$ is real, in Eqs. (\ref{10}), (\ref{11}) results in Eqs.
(\ref{5}) and (\ref{6}).

 Then, performing a corresponding change of
integration variables in Eq. (\ref{15}), and assuming that
\begin{eqnarray}
\beta_{R}=\frac{2 \pi}{a} \label{15.1.1}
\end{eqnarray}
in order to cover the whole Euclidian $(t_{E},\textbf{r}_{T},z)$
space, we get that the thermal partition function in the Rindler
wedge lying to one side of an infinite $xy$ plane, $Z_{R}$,  can be
expressed as the partition function of the Minkowski vacuum state,
$Z$,
\begin{eqnarray}
Tr[e^{- \frac{2 \pi}{a}H^{[\xi]}}]= \langle 0 | | 0 \rangle
\label{15.1}
\end{eqnarray}
with Unruh temperature that is proportional to a constant  proper
acceleration of an uniformly accelerated observer,
\begin{eqnarray}
T_{R}=\beta_{R}^{-1}=\frac{a}{2 \pi} . \label{15.1.1}
\end{eqnarray}
Equality (\ref{15.1}) means that  the expectation value in the
Minkowski vacuum state of any operators that are causally bounded
(restricted) to the (right) Rindler wedge is  equivalent to a
thermal average at the constant Unruh temperature \cite{Unruh-2}:
\begin{eqnarray}
\langle 0 |T_{t}(\phi (x_{1})\phi (x_{2})...) | 0 \rangle
=\frac{Tr[e^{- \beta_{R} H^{[\xi]}}T_{\xi}(\phi
(x_{1}(\xi_{1},\textbf{r}_{T1}, \eta_{1}))\phi
(x_{2}(\xi_{2},\textbf{r}_{T2},\eta_{2}))...)]}{Tr[e^{- \beta_{R}
H^{[\xi]}}]} \label{15.2}
\end{eqnarray}
where  $T_{t}$ and $T_{\xi}$ denote time and $\xi$ ordering,
respectively, and $x_{i}(\xi_{i},\textbf{r}_{Ti},\eta_{i})$
represents the same spacetime point as $x_{i}$ but in the Rindler
coordinates. Then  for an uniformly accelerated observer the
Minkowski vacuum is seen as  a thermal bath with temperature
proportional to the magnitude of the acceleration.

At this point one may wonder how an energy conservation is
maintained.  To see how it  proceeds, note that the left-hand-side
of Eq. (\ref{15.2}) is defined with respect to the Minkowski vacuum,
while the right-hand-side is defined with respect to the Rindler
vacuum.\footnote{The concept of vacuum is observer dependent: An
uniformly accelerated observer determines a zero energy state with
respect to $H^{[\xi]}$. } Their energy densities are different, as
one can see by calculating expectation values of the stress-tensor
$T_{\mu \nu}$ \cite{Milne}. Namely, if we accept that the energy
density of the Minkowski vacuum is equal to zero, then the energy
density of the Rindler vacuum is negative. The thermal state of the
Rindler quanta increases energy density from negative till zero
value and compensates this difference to maintain  energy
conservation.

Now, let us consider the future light cone and associate the Milne
frame with the system of the (hypothetical) observers which move
with different but constant longitudinal velocities in such a way
that their world lines begin at $z=t=0$.  In the Milne frame,
coordinates $(\xi, \eta)$ are related with Minkowski coordinates as:
\begin{eqnarray}
t'=b^{-1}e^{b\xi}\cosh b \eta, \label{17}
\\ z'=b^{-1}e^{b\xi}\sinh b
\eta , \label{18}
\end{eqnarray}
where primes are introduced to distinguish the  parametrization of
the Minkowski coordinates  in the future light cone from the
parametrization (\ref{10}), (\ref{11}) in the right Rindler wedge.
Here  $b$ is a scale-parameter, and the two other coordinates
$\textbf{r}_{T}=(x,y)$ are the same both in the Minkowski and Milne
frames. One can see  that the  $(- \infty, + \infty)$ region of
$\xi$, $\eta$ coordinates covers the whole future light cone. Taking
into account (\ref{17}) and (\ref{18}) we can calculate the
Minkowski line element restricted to the future light cone  and get
\begin{eqnarray}
ds^{2}= dt'^{2}- d\textbf{r}_{T}^{2}- dz'^{2}=e^{2b\xi}( d\xi^{2}-d
\eta^{2}) - d\textbf{r}_{T}^{2}. \label{19}
\end{eqnarray}
It follows from Eq. (\ref{19}) that the Minkowski metrics is
non-static with respect to $\xi$. It is convenient to  introduce
dimensionless variables $\overline{\eta}$, $\overline{\xi}$:
\begin{eqnarray}
\overline{\eta} = b \eta, \label{19.01}
\\ \overline{\xi} =b \xi , \label{19.02}
\end{eqnarray}
and define
\begin{eqnarray}
\tau = b^{-1}e^{\overline{\xi}}. \label{20}
\end{eqnarray}
Then
\begin{eqnarray}
t'=\tau \cosh \overline{\eta}, \label{21.1}\\ z'=\tau \sinh
\overline{\eta}, \label{21.2}
\end{eqnarray}
and one can see that $\tau$ is a proper time of an inertial Milne
observer with constant rapidity $\overline{\eta}$ and constant
transverse coordinates $\textbf{r}_{T}$, because
\begin{eqnarray}
ds^{2}=dt'^{2}-d\textbf{r}_{T}^{2}-dz'^{2}=  d\tau^{2}- \tau^{2}d
\overline{\eta}^{2} - d\textbf{r}_{T}^{2}. \label{22}
\end{eqnarray}
To define Lagrangian density in the future light cone in the
coordinates $(\xi, \eta)$, one can rewrite   action (\ref{1}) in the
Milne frame  (\ref{17}), (\ref{18}) as
\begin{eqnarray}
S=\int d\xi d\eta  d^{2}r_{T} \left [\frac{1}{2} \left
(\frac{\partial \phi}{\partial \xi}\right )^{2} - \frac{1}{2} \left
(\frac{\partial \phi}{\partial \eta}\right )^{2} - e^{2b\xi} \left
(\frac{1}{2}\left (\frac{\partial \phi}{\partial
\textbf{r}_{T}}\right )^{2}+ V(\phi)\right ) \right]\equiv
 \label{26.1} \\
\int d\xi d\eta d^{2}r_{T} L_{M}, \label{26.2}
\end{eqnarray}
where $L_{M}$ is the corresponding Lagrangian density, and we take
into account that
\begin{eqnarray}
dtdz=d\eta d \xi e^{2b\xi}. \label{19.1}
\end{eqnarray}
The  Hamiltonian density, $H_{M}$, is
\begin{eqnarray}
H_{M}= \frac{1}{2} \left (\frac{\partial \phi}{\partial \xi}\right
)^{2}+ \frac{1}{2} \left (\frac{\partial \phi}{\partial \eta}\right
)^{2} +  e^{2b\xi} \left (\frac{1}{2}\left (\frac{\partial
\phi}{\partial \textbf{r}_{T}}\right )^{2}+ V(\phi)\right ),
\label{28}
\end{eqnarray}
and the corresponding Hamiltonian that generates translation in the
time-like direction with respect to $\xi$ is $H^{[\xi]}=\int d\eta
d^{2}r_{T}H_{M}$. Notice that unlike the Hamiltonian of an uniformly
accelerated observer in the Rindler wedge, $H^{[\xi]}$ is explicitly
time-dependent ($\xi$ is the time-like parameter in the future light
cone). To proceed further, let us  assume that the term in the
Hamiltonian with explicit $\xi$-dependence is small and can be
neglected. To see under what conditions it is the case, let us
rewrite action (\ref{26.1}) in variables $(\tau, \textbf{r}_{T},
\overline{\eta})$, then
\begin{eqnarray}
S=\int d\tau d\overline{\eta } d^{2}r_{T} \left [\frac{\tau}{2}
\left (\frac{\partial \phi}{\partial \tau}\right )^{2} -
\frac{1}{2\tau} \left (\frac{\partial \phi}{\partial
\overline{\eta}}\right )^{2} - \tau \left (\frac{1}{2}\left
(\frac{\partial \phi}{\partial \textbf{r}_{T}}\right )^{2}+
V(\phi)\right ) \right]\equiv
\\ \label{26.01}
\int d\tau d\overline{\eta} d^{2}r_{T} \overline{L}_{M},
\label{26.002}
\end{eqnarray}
where $\overline{L}_{M}$ is the corresponding Lagrangian density.
The corresponding Hamiltonian density, $\overline{H}_{M}$, reads
\begin{eqnarray}
\overline{H}_{M}= \frac{\tau }{2}  \left (\frac{\partial
\phi}{\partial \tau}\right )^{2}+ \frac{1}{2\tau} \left
(\frac{\partial \phi}{\partial \overline{\eta}}\right )^{2} + \tau
\left (\frac{1}{2}\left (\frac{\partial \phi}{\partial
\textbf{r}_{T}}\right )^{2}+ V(\phi)\right ). \label{28.01}
\end{eqnarray}
The  Hamiltonian that generates translation in the time-like
direction with respect to $\tau$ is $H^{[\tau]}=\int
d\overline{\eta} d^{2}r_{T}\overline{H}_{M}$.  Note that
\begin{eqnarray}
H^{[\xi]}= b \tau H^{[\tau]}, \label{28.02}
\end{eqnarray}
and that the dependence on the scale-parameter $b$ is canceled out
in $H^{[\tau]}$. It follows from (\ref{28.01})  that the last term
in $\overline{H}_{M}$  is small and can be neglected  if
interactions are weak,  and if the  spacetime is effectively
two-dimensional, the latter means that only modes with low
transverse momenta, $p_{T}^{2} \tau^{2} \ll 1$, are
considered.\footnote{Also, one can notice from Eq. (\ref{28.01})
that massive quantum field is effectively massless for very early
times when $m \tau \ll 1$.}

To find an expression for the vacuum partition function $Z= \langle
0 | | 0 \rangle $  in the future light cone, let us assume that the
above mentioned conditions are satisfied, and hypothesize that the
vacuum partition function, restricted on the future light cone, can
be associated in the weak coupling approximation and for low $p_{T}$
particles with the thermal partition function $Z_{M} $,
\begin{eqnarray}
 Z_{M}=Tr[e^{- \beta_{M} H^{[\xi]}}]= Tr[e^{- \beta_{M} b \tau H^{[\tau]}}], \label{29.01}
\end{eqnarray}
where $\beta_{M}$ will be specified below.

To establish a relation in the future light cone between the vacuum
partition function $Z= \langle 0 | | 0 \rangle $ and the thermal
partition function $Z_{M}$, approximate first  $Z_{M}$  as
\begin{eqnarray}
 Z_{M}\simeq Z_{FM}= Tr[e^{- \beta_{M} H^{[\xi]}_{F}}], \label{29.1}
\end{eqnarray}
where $H^{[\xi]}_{F}=\int d \eta d^{2}r_{T} H_{FM}$, and $H_{FM}$ is
the Hamiltonian density of free massless field in the $(\xi, \eta)$
spacetime,
\begin{eqnarray}
H_{FM} = \frac{1}{2} \left (\frac{\partial \phi}{\partial \xi}\right
)^{2}+ \frac{1}{2} \left (\frac{\partial \phi}{\partial \eta}\right
)^{2}. \label{28.1}
\end{eqnarray}
Then,   let us rewrite   $Z_{FM}$  as the Euclidian functional
integral,
\begin{eqnarray}
Z_{FM}=N_{FE} \int D \phi \exp{\left [   \int_{0}^{\beta_{M}} d
\zeta \int d^{2}r_{T}  d \eta L_{FM} (\phi, i\frac{\partial
\phi}{\partial
\zeta} )\right ]} =  \nonumber \\
N_{FE} \int D \phi \exp{\left [ - \int_{0}^{\beta_{M}} d \zeta \int
d^{2}r_{T}  d \eta \left (\frac{1}{2} \left (\frac{\partial
\phi}{\partial \zeta}\right )^{2} + \frac{1}{2} \left
(\frac{\partial \phi}{\partial \eta}\right )^{2}\right )\right ]}
\label{30}
\end{eqnarray}
with periodic boundary conditions $\phi(\zeta =0) = \phi(\zeta
=\beta_{M})$  for arbitrary $\beta_{M}$ with respect to  $\zeta$,
here $\zeta$ is real and $\xi = i \zeta$.

Now, notice that  for free massless field in effectively
two-dimensional spacetime the functional form of the Euclidian
functional integral (\ref{15}) in the Rindler wedge non-inertial
coordinate system is
\begin{eqnarray}
Z_{FR}= N_{FE} \int D \phi \exp{\left [  - \int_{0}^{\beta_{R}} d
\zeta \int d^{2}r_{T}  d \eta \left (\frac{1}{2} \left
(\frac{\partial \phi}{\partial \zeta}\right )^{2} + \frac{1}{2}
\left (\frac{\partial \phi}{\partial \eta}\right )^{2} \right)\right
]}, \label{30.1}
\end{eqnarray}
and thereby is identical with functional form of the Euclidian
functional integral (\ref{30}) in the future light cone inertial
coordinate system. Then, to establish relation of $Z_{FM}$ with the
Minkowski vacuum partition function, one can apply the same method
which is used for the thermal partition function $Z_{R}$ in the
Rindler wedge. Namely,  assuming that $\beta_{M} = 2 \pi /b $ and
changing integration variables\footnote{It is worth emphasizing, to
avoid misunderstanding, that this is just a change of integration
variables but not coordinates. }
\begin{eqnarray}
t_{E}=b^{-1}e^{b\eta}\sin b\zeta  , \label{5-1}\\
z =b^{-1}e^{b\eta}\cos b \zeta , \label{6-1}
\end{eqnarray}
in Eq. (\ref{30}), we get that
\begin{eqnarray}
 Z_{FM}=  \langle 0 | | 0 \rangle  \label{30.2}
\end{eqnarray}
for $\beta_{M} = 2 \pi /b $. Finally, assuming that $ Z_{M}\simeq
Z_{FM}$ and  taking into account Eq. (\ref{29.01}), we get that with
respect to $H^{[\tau]}$ the pure Minkowski vacuum state looks like
mixed thermal state with temperature $1/2 \pi \tau$,
\begin{eqnarray}
 \langle 0 | | 0 \rangle \simeq Tr[e^{- 2 \pi \tau H^{[\tau]}}].
 \label{29.1}
\end{eqnarray}
Notice that  the value of the auxiliary scale-parameter $b$ is
arbitrary: correspondence with the Unruh effect is reached for any
$b$, and  the dependence on the scale-parameter $b$ is canceled out
in variables $(\tau, \textbf{r}_{T}, \overline{\eta})$. Equality
(\ref{29.1}) means that  the expectation value in the Minkowski
vacuum state of  quantum operators depending on the values of the
quantum field in spacetime points belonging only to the region
inside the future light cone is approximately  equivalent to a
thermal average at the temperature $1/2\pi \tau$.

The corresponding thermal statistical operator in the
right-hand-side of Eq. (\ref{29.1}) is defined over the analog of
the Rindler vacuum, that is one of the states in the expanding Milne
universe \cite{Milne,Unruh-rev}. It is known  that the difference
between expectation values of the energy-momentum tensor of massless
fields calculated in the analog of the Rindler vacuum and in the
Minkowski vacuum is negative, time-dependent and tends to zero for
asymptotic times \cite{Milne}. Then, if we accept that the energy
density of the Minkowski vacuum is equal to zero, the energy density
of the analog of the Rindler vacuum is negative, and, by
correspondence to the Unruh effect, the thermal state in the
right-hand-side of Eq. (\ref{29.1}) increases energy density from
negative till zero value.

Finally, let us consider our findings in view of relativistic heavy
ion collisions and  estimate the relevant parameters. First, recall
that in the  weak coupling limit of the QCD (corresponding to the
high energy limit  $\sqrt{s} \rightarrow \infty$ of collisions of
heavy nuclei) the initial conditions of nuclear collisions are
fairly well understood in terms of the Color Glass Condensate
framework \cite{CGC-1}, that is the effective field theory which
describes universal properties of saturated gluons in  wave
functions of colliding nuclei.  Saturation is characterized by a
transverse momentum scale $Q_{s}$, typical values are estimated to
be $Q_{s}^{2}\simeq 2$ GeV$^2$ at RHIC and $Q_{s}^{2}\simeq 5$
GeV$^2$ at the LHC \cite{CGC-2}. Then, in the idealized high-energy
limit of heavy ion collisions, the dynamics of the system right
after the collision  (at $\tau \ll Q_{s}^{-1}$) is that of
over-occupied far-from-equilibrium gluon fields expanding in the
longitudinal direction, with typical momentum $Q_s$ and a weak gauge
coupling $\alpha_s (Q_{s})$, usually referred to as the  Glasma
\cite{Glasma}. Because of the over-occupation, the system is
initially strongly interacting even though the coupling is weak.  At
very early times $\tau \ll Q_{s}^{-1}$, the dynamics of the
nonequilibrium Glasma created in such a collision  is described in
the midrapidity region with  approximately boost invariant classical
gauge fields screened on transverse distance scales $1/Q_{s}$,
rather than with particles (gluons). The classical fields are
decayed (i.e., field expectation value becomes to be zero) and
gluons are freed in a timescale $\tau \sim Q_{s}^{-1}$, see e.g.
Refs. \cite{Berges-a,Berges-b}. One can expect that our results can
be applicable after the decay of the classical gluon fields, i.e.,
at a lower bound of about $\tau_{in} \simeq Q_{s}^{-1}$, when
dynamics becomes governed by the quantum vacuum and its excitations.
It corresponds to a temperature $T_{in} =
 1/2 \pi \tau_{in}  \simeq Q_{s}/2 \pi  $ of the Minkowski vacuum thermal bath.\footnote{It is worth noting  that the same expression
 for temperature was proposed in Refs. \cite{Kharzeev-1-a,Kharzeev-1-b,Kharzeev-1-c} for partons moving in strong color field with typical
 (transient) acceleration $\sim Q_{s}$ in analogy to the Unruh effect. Notice,
 however, that our approach  here is quite different and is not based on a picture of accelerated
partons at all.} At the LHC, one can estimate $\tau_{in}\simeq 0.1$
fm and  $T_{in} \simeq 0.35$ GeV. Notice here, to avoid
misunderstanding, that the latter is not the temperature of the
whole quark-gluon system at $\tau_{in}\simeq 0.1$ fm;  in fact,
$T_{in} \simeq 0.35$ GeV is the temperature of the Minkowski vacuum
thermal bath only.

Thermalization mechanism in relativistic heavy ion collisions is not
yet fully understood. In the bottom-up thermalization scenario
\cite{bottom}, for example, the  pre-equilibrium evolution after
decay of the classical color fields is divided into three temporal
stages: (i) the system is dominated by the over-occupied hard gluons
whose typical transverse momentum is $Q_s$; (ii) soft gluons are
produced by collinear splitting processes; (iii) soft gluons
thermalize first and form the thermal bath, then the thermal bath
drains the energy from the hard gluons and make them thermalized.
Therefore, in such a scenario thermalization proceeds from bottom to
top in the energy scale. The approach which takes into account
thermal-like properties of the Minkowski vacuum in the future light
cone of relativistic heavy ion collisions suggests modification of
the bottom-up thermalization scenario. Namely, because the hard
gluons whose typical transverse momentum is $Q_s$  are immersed in
the Minkowski vacuum thermal bath with the temperature $T=1/2 \pi
\tau$, they start to thermalize already at the stage (i),
immediately after they are freed from classical fields at $\tau
\simeq Q_{s}^{-1}$.

Evidently,  in the course of the system evolution  weak coupling
approximation becomes not valid. Then, strictly speaking, Eq.
(\ref{29.1}) is applicable for very early proper times only.
However, even at later proper times one can expect that the pure
Minkowski vacuum state looks like a mixed state with some
thermal-like properties with respect to $H^{[\tau]}$.  With increase
of $\tau$ the differences between the pure ground (vacuum) state of
the Hamiltonian $H^{[\tau]}$ and the Minkowski vacuum state decrease
resulting in gradual disappearance of the thermal-like properties of
the Minkowski vacuum  with respect to the generator of the time-like
translations in the future light cone, $H^{[\tau]}$.

\section{Conclusions}

We have studied how  the global pure vacuum state in the Minkowski
spacetime looks like for the Milne set of inertial observers that
are locally restricted to the light cone with beginning at $t=z=0$.
We found that in a weak coupling approximation of effectively
two-dimensional massless scalar field model a pure Minkowski vacuum
state looks like  the mixed thermal state with  the Hamiltonian
$H^{[\tau]}$ which is the generator of translations in the time-like
direction with respect to (longitudinal) proper time  $\tau =
\sqrt{t^{2} - z^{2}}$, and whose lowest energy eigenstate does not
coincide with the Minkowski vacuum. Effective spacetime
two-dimensionality for such a system   means that we consider only
$p_{T} \tau \ll 1 $ modes (particles). We found that temperature of
the corresponding thermal state varies with respect to  proper time
$\tau$  as $1/2 \pi\tau$. In other words, the Minkowski vacuum
expectation value of the appropriate local quantum operators  in the
light cone can be interpreted in terms of a thermal-like mixture of
states which differ locally from the Minkowski vacuum, an analogue
of the Unruh effect.

In  a relativistic nucleus-nucleus collision the abstract Milne
coordinate system becomes physical reference frame, and our analysis
suggests that created at $\tau \simeq Q_{s}^{-1}$ in the future
light cone of  a relativistic heavy ion collision quarks and gluons
feel quantum fluctuations of the Minkowski vacuum as a quantum
thermal bath of weakly interacting on-mass-shell soft gluons. That
would speed up the process of ``hydrodynamization'' and, perhaps,
can be responsible  for the ``direct photon flow puzzle''  observed
in ultrarelativistic heavy ion collisions  \cite{photon-a,photon-b}.

\begin{acknowledgments}
I am grateful to J. Berges,  K. Boguslavski, and S. Floerchinger for
discussions. This research was supported in part by the ExtreMe
Matter Institute EMMI   at the GSI Helmholtzzentrum f\"ur
Schwerionenforschung, Darmstadt, Germany.
\end{acknowledgments}

\end{document}